\documentclass[twocolumn]{aastex62}
\pdfoutput=1 
\usepackage{amsmath,amstext}
\usepackage{apjfonts} 
\usepackage[figure,figure*]{hypcap}
\usepackage{graphicx}	
\usepackage{amssymb}	
\usepackage{booktabs}
\usepackage{chemfig}

\received{2018 April 4}
\revised{2018 May 3}
\accepted{2018 May 14}

\begin{document}

\title{Evidence for multiple molecular species in the hot Jupiter HD~209458b}

\correspondingauthor{Nikku Madhusudhan}
\email{nmadhu@ast.cam.ac.uk}

\author{George A. Hawker}
\affiliation{Institute of Astronomy, University of Cambridge, Madingley Road, Cambridge, CB3 0HA, UK}
\author{Nikku Madhusudhan}
\affiliation{Institute of Astronomy, University of Cambridge, Madingley Road, Cambridge, CB3 0HA, UK}
\author{Samuel H. C. Cabot}
\affiliation{Institute of Astronomy, University of Cambridge, Madingley Road, Cambridge, CB3 0HA, UK}
\author{Siddharth Gandhi}
\affiliation{Institute of Astronomy, University of Cambridge, Madingley Road, Cambridge, CB3 0HA, UK}
\begin{abstract}
Molecular species in planetary atmospheres provide key insights into their atmospheric processes and formation conditions. In recent years, high-resolution Doppler spectroscopy in the near-infrared has allowed detections of H$_2$O and CO in the atmospheres of several hot Jupiters. This method involves monitoring the spectral lines of the planetary thermal emission Doppler-shifted due to the radial velocity of the planet over its orbit. However, aside from CO and H$_2$O, which are the primary oxygen- and carbon-bearing species in hot H$_2$-rich atmospheres, little else is known about molecular compositions of hot Jupiters. Several recent studies have suggested the importance and detectability of nitrogen-bearing species in such atmospheres. In this Letter, we confirm potential detections of CO and H$_2$O in the hot Jupiter HD 209458b using high-resolution spectroscopy. We also report a cross-correlation peak with a signal-to-noise ratio of $4.7$ from a search for HCN. The results are obtained using high-resolution phase-resolved spectroscopy with the Very Large telescope CRyogenic high-resolution InfraRed Echelle Spectrograph (VLT CRIRES) and standard analysis methods reported in the literature. A more robust treatment of telluric contamination and other residuals would improve confidence and enable unambiguous molecular detections. The presence of HCN could provide constraints on the C/O ratio of HD~209458b and its potential origins.

\end{abstract}

\section{Introduction} \label{sec:intro}
Exoplanets orbiting nearby sun-like stars provide a unique opportunity to study the diversity of planetary processes that can arise from primordial environments like our own, particularly through chemical species in their atmospheres. Due to the low temperatures ($\lesssim$ 100~K) of solar system giant planets, several important chemical tracers of planetary origins are either condensed out, such as H$_2$O \citep{wong2004, atreya2016}, or present in trace quantities such as CO and HCN, making their origins ambiguous \citep{moreno2003, cavalie2008, cavalie2010, moses2010}. It has been predicted that atmospheres of hot giant exoplanets should contain copious amounts of these molecules and can, therefore, provide critical insights into planetary formation \citep{madhu2016}. 

\begin{figure*}[ht!]
\centering
\includegraphics[width=150mm,trim={10cm 15cm 10cm 15cm},clip]{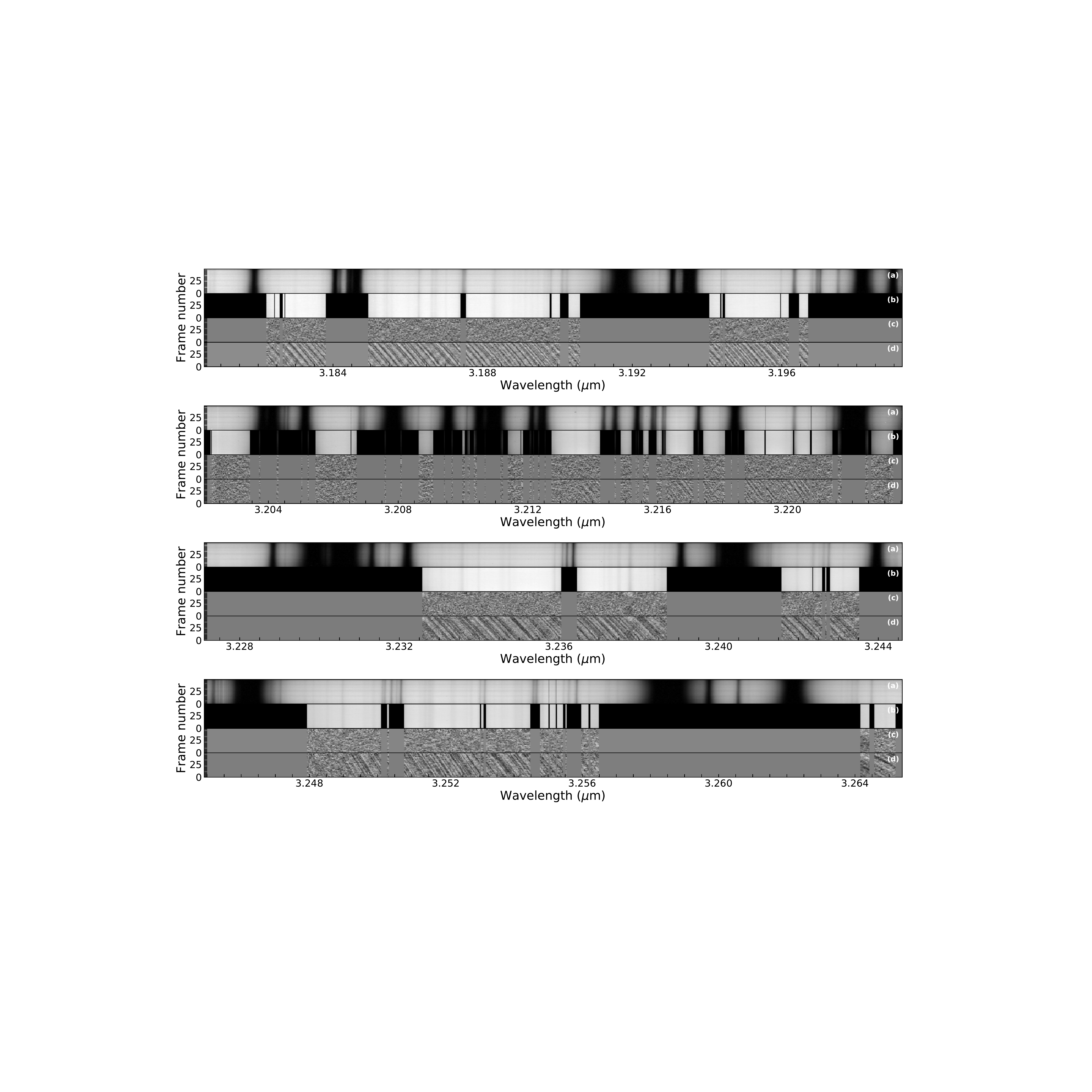}
\caption{Stages of detrending, for detectors 1--4 and the set of observations taken on 2011 July 25. The x-axis corresponds to wavelength, and the y-axis corresponds to frame number, increasing in time. Panel (a): spectra immediately after reduction of nodding frames. Heavy telluric contamination is evident (e.g. at 3.1915 $\mu$m). 
Poor seeing conditions manifest as dark horizontal bands. Panel (b): reduced spectra after wavelength calibration, alignment, additional cleaning, normalization, and masking. This image (excluding masked regions) is the input of our detrending algorithm. Panel (c): data subject to column-wise mean subtraction, the optimal number of SYSREM iterations, a 15-pixel standard-deviation high-pass filter, and column-wise standard-deviation division. Panel (d): the same as in Panel (c), but with the injection of our planet model at 40x its nominal strength prior to detrending. The preserved planetary absorption features appear as dark trails that stretch over $\sim0.0008 \: \mu$m.}
\label{fig:detrendinghcn2}
\end{figure*}

The hot Jupiter HD~209458b is one of the most favorable targets for atmospheric characterization. The planet orbits a bright ($V$=7.65) sun-like (G0V) star in a 3.5 day period and has mass 0.69$\pm$0.017 $M_J$ and radius 1.38$\pm$0.018 $R_J$ \citep{knutson2007}. However, with its close-in orbit and mass between those of Saturn and Jupiter, the planet has no analog in the solar system, even though it likely originated in similar conditions to the primordial solar nebula given its solar-like host star. Low-resolution spectra of the planet obtained using the \textit{Hubble Space Telescope} (HST) revealed the presence of H$_2$O in its atmosphere, albeit with significantly weaker spectral features than originally anticipated \citep{deming2013, line2016}. While the H$_2$O abundance has been found to be significantly sub-solar at the day--night terminator region of the atmosphere \citep{madhu2014b, barstow2017,macdonald2017a}, the same is not well constrained in the dayside atmosphere \citep{line2016}. Efforts to detect any other molecule using low-resolution spectra of the planet have proved elusive to date. Recently, an HST transmission spectrum of the planet initially suggested strong evidence for NH$_3$ and/or HCN at the terminator \citep{macdonald2017a}. However, a subsequent study including additional data and modeling lowered the detection significances, leaving HCN undetected \citep{macdonald2017}. 

In recent years, high-resolution Doppler spectroscopy has enabled detections of key molecules in the atmospheres of hot Jupiters \citep{snellen2010,brogi2012}. This method involves monitoring numerous (10$^2$-10$^3$) individual molecular lines in the planetary spectrum being Doppler shifted as the planet traverses its orbit, leading to a high-fidelity detection of the molecule. Such observations have led to detections of H$_2$O and CO in several hot Jupiters \citep{snellen2010, brogi2012, birkby2013, rodler2013, birkby2017}, and TiO in one \citep{nugroho2017}. The previous application of this method to HD 209458b has led to the detection of CO at the day--night terminator region of its atmosphere  \citep{snellen2010} as well as both CO and possibly H$_2$O in the dayside \citep{brogi2017}. In this Letter we use this method in search of molecular signatures in the dayside atmosphere of the planet. 

\section{Observations and Reduction} \label{sec:obs}
We obtain spectroscopic observations of the system observed with the CRyogenic high-resolution InfraRed Echelle Spectrograph (CRIRES; \citealt{kaeufl2004}) on the Very Large Telescope (VLT), in Chile, made available through the European Southern Observatory (ESO) Science Archive.

\subsection{Observations}

The data were obtained as nodding frames across four nights (2011 July 18, 25, August 4, September 6) covering two spectral ranges (2.29 - 2.35 $\mu$m and 3.18 - 3.27 $\mu$m) as part of the CRIRES survey of hot Jupiter atmospheres \citep{2011IAUS..276..208S}. The instrument setup included a 0.2" slit for the highest possible wavelength resolution of $R\sim$10$^5$, and made use of the Multi Applications Curvature Adaptive Optics (MACAO). The spectrograph contains four (512 $\times$ 1024 pixel) Aladdin III InSb detectors, separated by gaps of 280 pixels. The telescope is nodded along the slit direction by 10" in a typical ABBA sequence; pairs of exposures taken at different nodding positions may be combined to accurately remove the background and improve the signal. The observations target bright, dayside emission from the companion planet HD 209458b, near secondary eclipse. Each set of spectra are phase resolved in the ranges $\phi\sim$ 0.5 - 0.6 and $\phi\sim$ 0.4 - 0.5 for the 2.3$\mu$m and 3.2$\mu$m bands, respectively. The planet has an orbital period $T$ = 3.52 days and reference crossing time $t_{\phi}$=0 = 52854.325456 JD \citep{butler2006}. 

\subsection{Initial Reduction}
Data from each detector on each observation night are treated separately throughout the following procedures. We use \texttt{Esorex} from the ESO CRIRES reduction toolkit (v2.3.4) to reduce the two-dimensional spectral images, perform flat-fielding and bad pixel corrections, and optimally extract \citep{horne1986} one-dimensional spectra. We perform additional calibrations with our custom pipeline, X-COR, written in Python 2.7. Remaining bad pixels and regions are replaced with the linear interpolation of their neighbors. Next, all of spectra are aligned to the highest signal-to-noise ratio spectrum. The shifts are typically sub-pixel, and account for a small drift in wavelength alignment throughout the night. We determine a wavelength solution by matching strong, telluric absorption lines in the data to those in an ATRAN \citep{lord1992} synthetic transmission spectrum, and fitting the offsets with a cubic polynomial. The centroids are aligned with an estimated uncertainty of $\sim$0.5 km s$^{-1}$. Seeing variations throughout the night cause the spectra to have different baseline continua. We model this variation by averaging over the brightest pixels in each spectrum, and remove it through division. We arrange our calibrated data in an $N$ $\times$ 1024 array, where $N$ is the number of reduced spectra for a given night. The x-axis corresponds to the detector wavelength bins, and the y-axis corresponds to orbital phase or time. We apply a mask to remaining bad pixels and telluric zones identified as high variance or low mean columns, along with the ends of the array to avoid edge artifacts. 

\begin{figure}[ht!]
\centering
\includegraphics[width=1\linewidth]{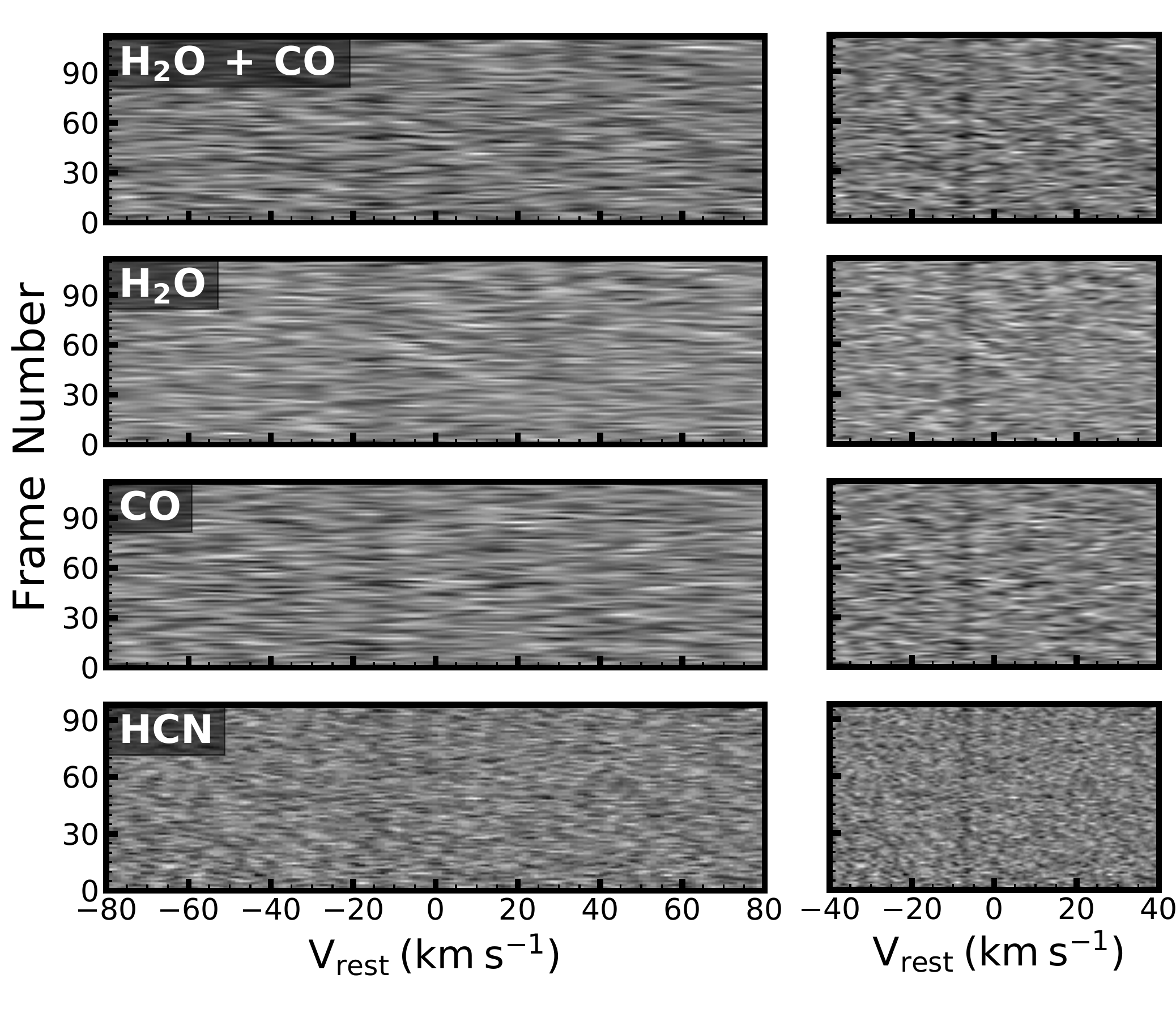}

\caption{Frame-by-frame cross correlation values as a function of velocity. The CCFs are shifted into the planetary rest frame by the peak significance $K_p$, for each model template (HCN, H$_2$O, CO, and their combination). The right panel shows the CCF with a model injection at two times nominal strength. Each CCF contains a dark vertical trail at approximately the known systemic velocity of $-$14.8 km s$^{-1}$ from alignment between the model template and intrinsic features.}
\label{fig:ccf_plot}
\end{figure}

\subsection{Detrending with SYSREM} 

\begin{figure*}[ht!]
\centering
	\includegraphics[width=130mm,trim={2cm 0 4cm 0},clip]{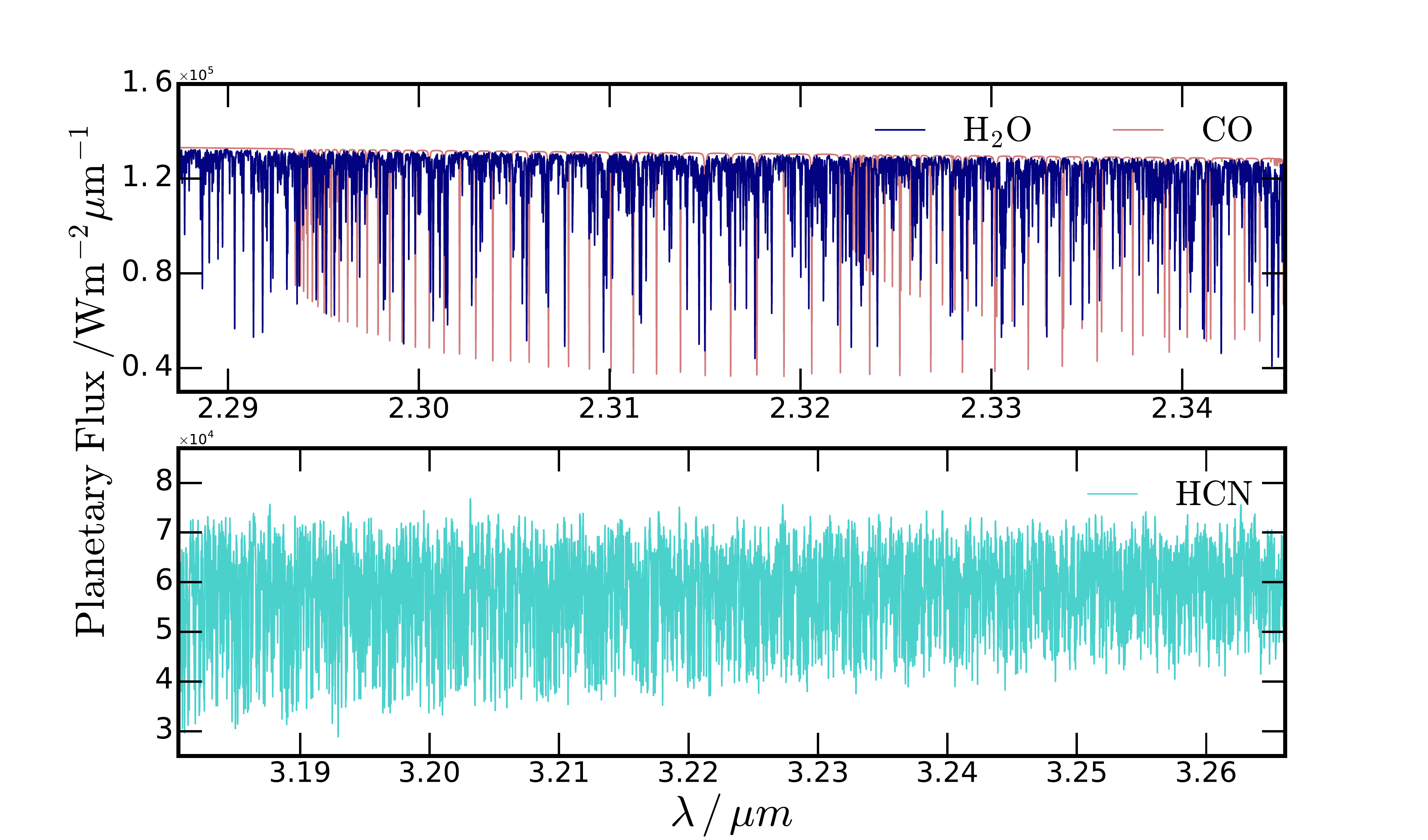}
    \caption{High-resolution model spectra of HD~209458b showing line features of HCN, H$_2$O, and CO in the observed bands at 2.28-2.35$\mu$m and 3.18-3.27$\mu$m.}     
\label{fig:flux}
\end{figure*}

The stellar signal and telluric absorption caused by Earth's atmosphere are approximately time-invariant, whereas the planetary spectrum is subjected to a considerable Doppler shift ($\sim34$ pixels) over the course of the night. Following similar studies \citep{birkby2013,birkby2017,nugroho2017} we adopt SYSREM \citep{tamuz_2005}, an algorithm based on principal component subtraction that weights the data by its uncertainty (the sum of spectral extraction error and shot-noise added in quadrature). SYSREM iterations progressively remove the stellar signal, telluric lines, and trends from environmental effects (e.g. airmass), and eventually target sub-pixel commonalities in the planet signal. As in previous studies \citep{birkby2013,birkby2017,nugroho2017}, we calibrate the number of iterations to use in our analysis by injecting our model planetary spectrum at the expected $K_\mathrm{p}$ (the orbital velocity semi-amplitude) and $V_\mathrm{sys}$ (systemic velocity) and maximizing the significance of its recovery. Factors such as tellurics, the number of absorption lines within the detector's coverage, and systematics such as the odd-even effect influence the optimal number of iterations. Therefore, we optimize each detector in each wavelength band individually, and repeat for each molecular species. For most detectors, 2-6 iterations are optimal; however, heavy telluric contamination can require up to 13 iterations. We use a high-pass filter on each detrended spectrum to remove broadband variation, and normalize wavelength bins by their temporal standard deviation. Figure \ref{fig:detrendinghcn2} panel (a) shows the extracted spectra from 2011 July 25 in the 3.2$\mu$m spectral band. Panel (b) shows the normalised data with masking, panel (c) depicts the detrended data, and panel (d) shows the data with a 40$\times$ model injection prior to detrending.
\newline\newline
\subsection{Cross Correlation} \label{sec:cross-correlation}
We perform a cross-correlation analysis that scans the $K_p$--$V_{sys}$ parameter space in search of the planet signal. Following previous methods \citep{brogi2012}, we cross correlate using spectral templates consisting of narrow Gaussian profiles fit to the strongest absorption lines in the atmospheric model, discussed in Section~\ref{sec:models}. 

We Doppler shift the template by velocities ranging from -260 km s$^{-1}$ to 250 km s$^{-1}$ in steps of 1.0 km s$^{-1}$. At each velocity, we linearly interpolate the template onto the detector wavelength grid, and take the dot product with each detrended spectrum. We then combine the cross-correlation functions (CCFs) from all four detectors and both nights.  The resultant matrices are shown in Figure~\ref{fig:ccf_plot}. The dark trail at the planetary $V_{sys}$ represents cross correlation from the alignment of model and intrinsic features.

We sample $K_p$, the semi-amplitude of the orbital velocity, over a grid ranging from 30 km s$^{-1}$ to 250 km s$^{-1}$ in steps of 1.0 km s$^{-1}$, and shift each row of the CCF matrix by the orbital velocity $K_p\sin(2\pi\phi(t))$. For the correct $K_p$ this transformation shifts the CCF matrix into the rest frame of the star--planet system, which itself has the systemic velocity $V_{sys}$. We then sum the pixels within an $N$ $\times$ 5 sliding window, centered on velocities ranging from $-$80 km s$^{-1}$ to 80 km s$^{-1}$; this step samples potential values for $V_{sys}$. If the template absorption lines are present in the data, then we expect the sum to be maximized at the planetary $V_{sys}$. Finally the signal-to-noise ratio (S/N) is obtained by normalizing the cross-correlation sum by the standard deviation across the entire $K_p$-$V_{sys}$ range; this S/N metric represents the strength of the cross-correlation sum relative to noise. Figure \ref{fig:ccf_plot} shows the combined CCFs in the rest frame of the maximal $K_p$ both for the data (left) and with a model injection at two times nominal strength (right).
\vspace{5mm}

\begin{figure}[ht!]
\centering  
  \includegraphics[width=\linewidth]{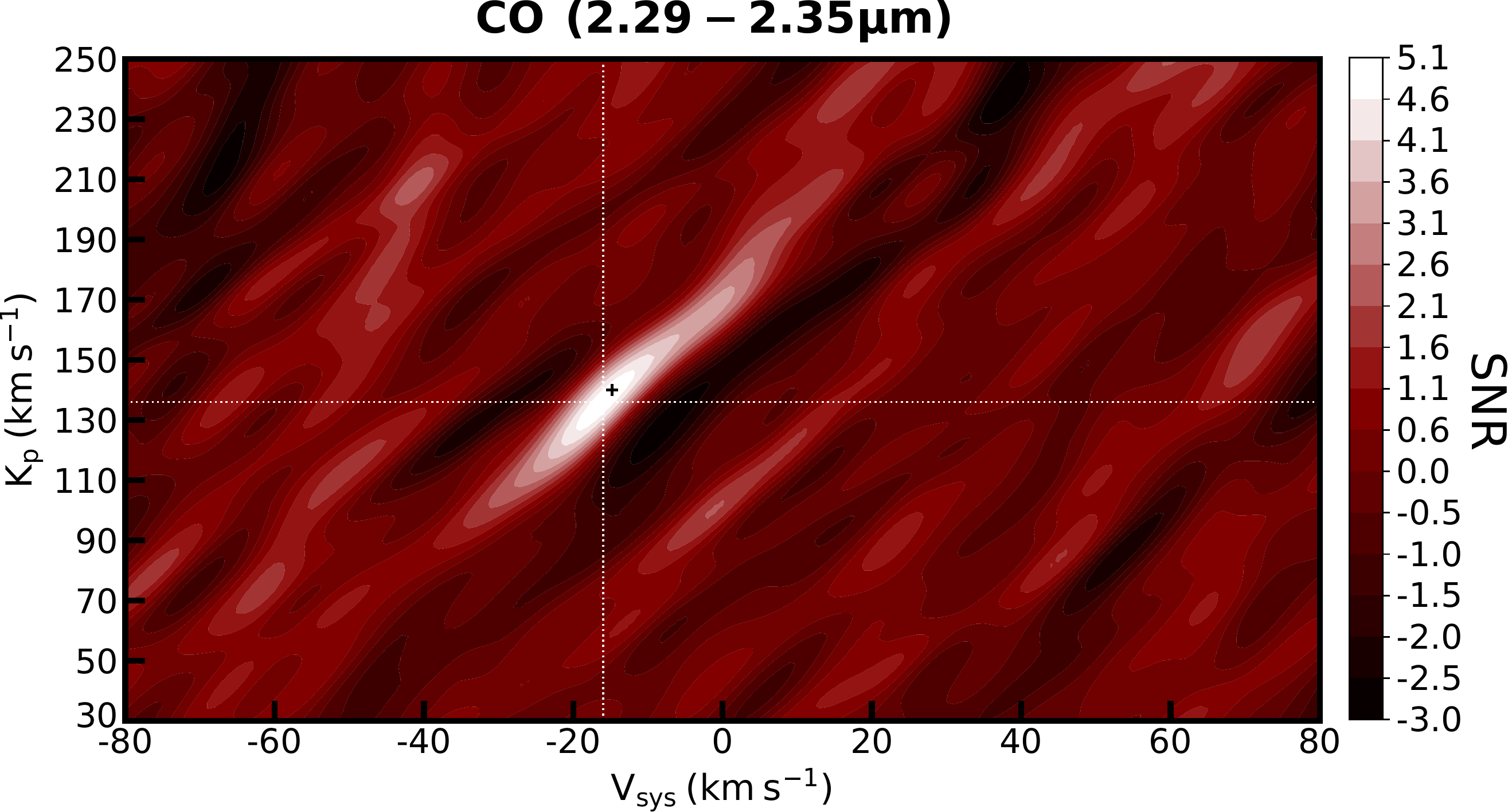}
  
  \vspace{0.1cm}
  \includegraphics[width=\linewidth]{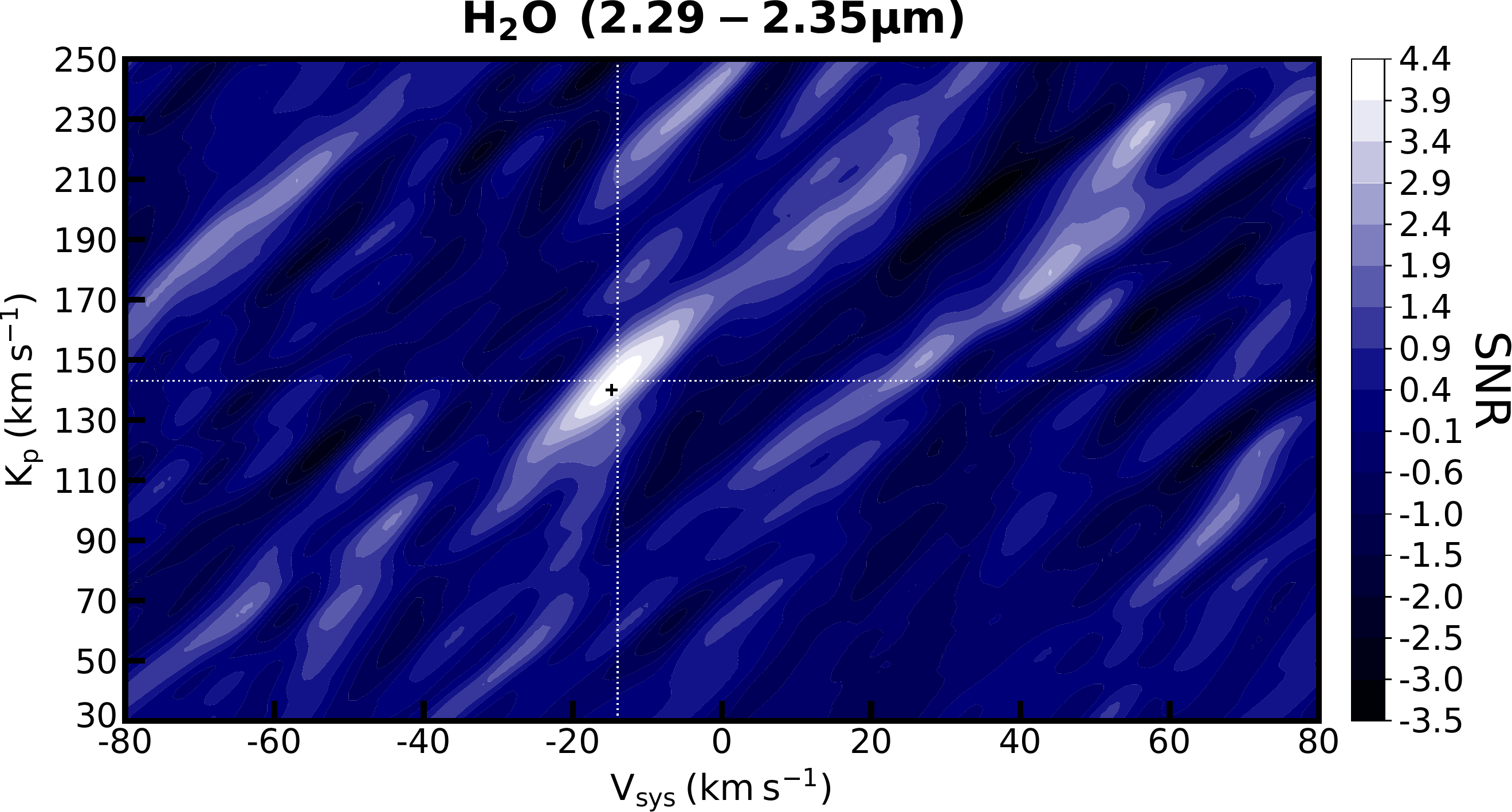}
  
  \vspace{0.1cm}
  \includegraphics[width=\linewidth]{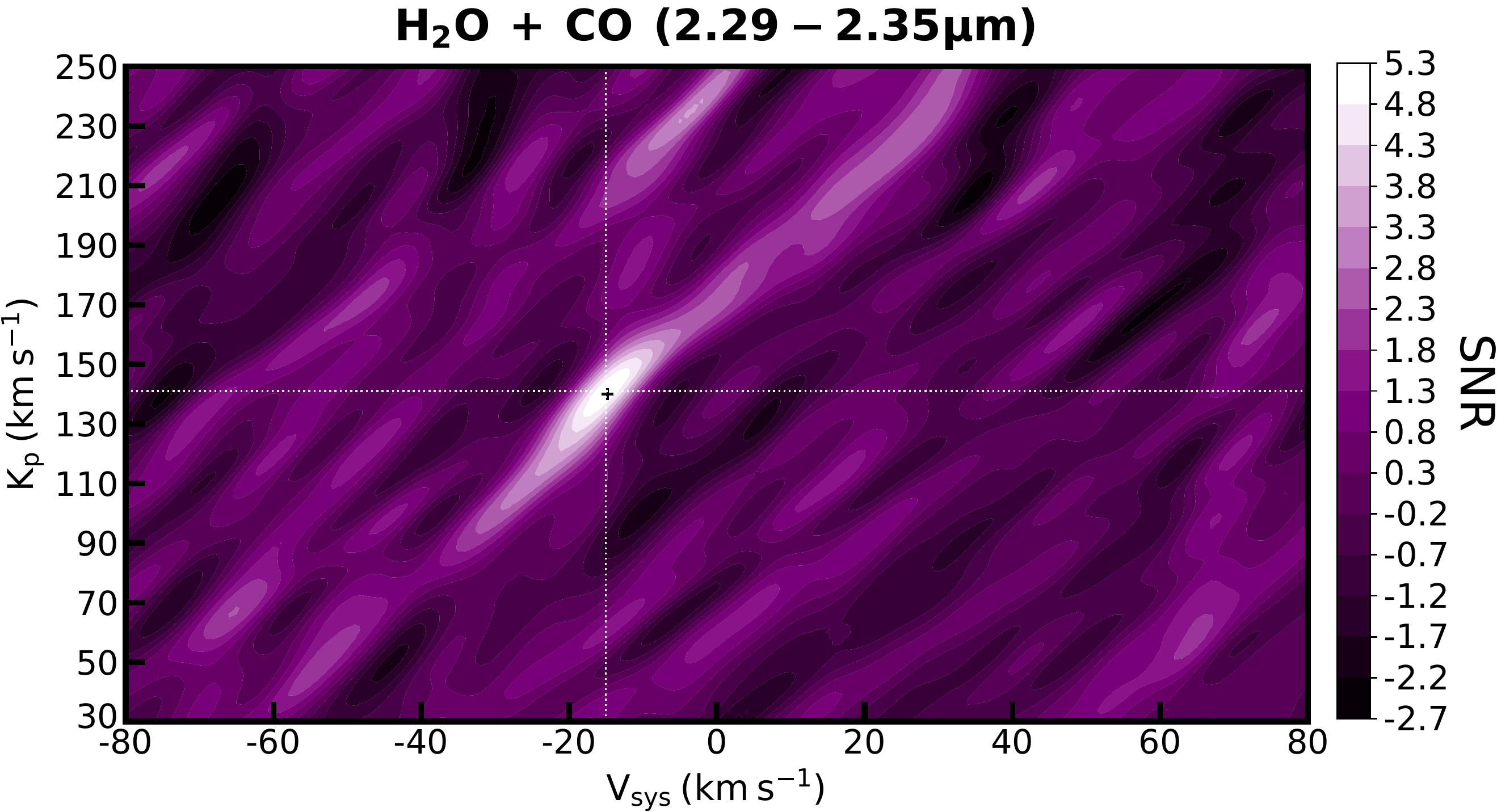}
  
  \vspace{0.1cm}
  \includegraphics[width=\linewidth]{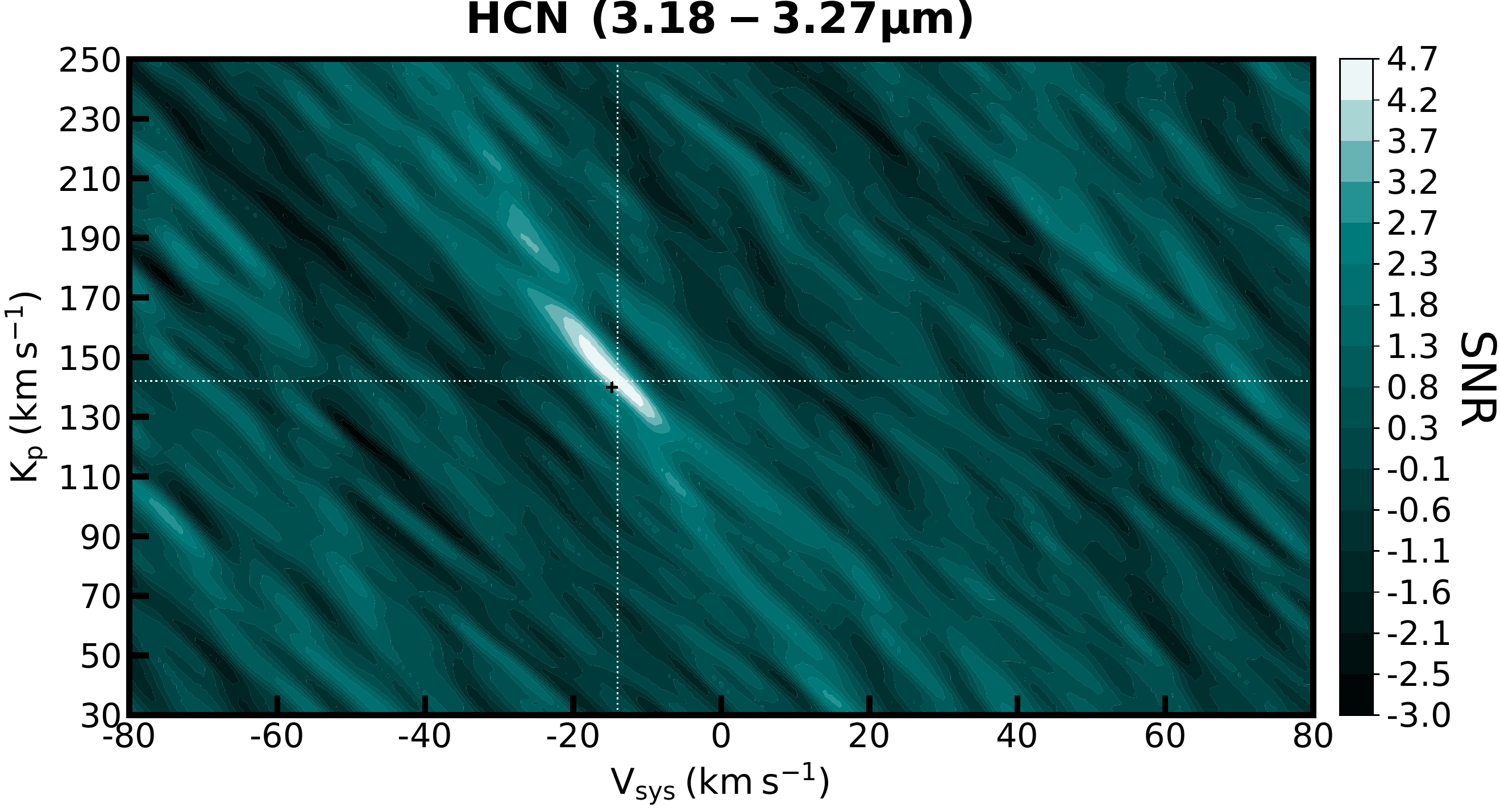}

\caption{
Cross correlation S/N for CO, H$_{2}$O, and HCN as a function of systemic velocity and peak radial velocity of the planet. Cross correlation in the 2.29-2.35$\mu$m spectral band yields an S/N of 5.1 with the CO template, and 4.4 with H$_{2}$O individually. A combined CO + H$_{2}$O model yields a boosted S/N of 5.3. HCN yields an S/N of 4.7 in the 3.18-3.27$\mu$m spectral band. The white crosshairs denote the $K_p$ and $V_{sys}$ for the peak S/N which agree with the black cross corresponding to known orbital parameters.}
\label{fig:test}
\end{figure}

\section{Modeling} \label{sec:models}

In order to cross correlate with the data we generate high-resolution model atmospheric spectra for HD 209458b using the GENESIS code \citep{gandhi_2017}. The models involve line-by-line radiative transfer computed via the Feautrier method to obtain the emergent spectra in the observed bands at a spectral resolution of $R\gtrsim$300,000 (Figure \ref{fig:flux}). The models assume a plane-parallel atmosphere in hydrostatic equilibrium and local thermodynamic equilibrium. The atmosphere comprises 150 model layers evenly spaced in $\log(P)$ for pressures from 10$^{-8}$-100 bar. We adopt the P--T profile derived from recent retrievals of dayside emission spectra both at low resolution \citep{line2016} as well as at high resolution \citep{brogi2017}. We also test a number of P--T profiles to confirm the robustness of our results, including a radiative-convective equilibrium profile \citep{gandhi_2017}. We consider molecular opacity due to the prominent sources of absorption in the observed bands. These include line absorption due to the molecules H$_2$O, CO, and HCN, and continuum opacity from collision-induced absorption (CIA) due to H$_2$--H$_2$ and H$_2$--He interactions \citep{richard_2012}. H$_2$O, CO, and HCN are known to be amongst the spectroscopically strong species containing oxygen, carbon, and nitrogen in the high-temperature H$_2$-rich atmospheres of hot Jupiters \citep{madhu2012,moses2013}. These molecules are also particularly conducive to high-resolution spectroscopy given their large number of rovibrational transitions in the near-infrared. 

We use the latest high-temperature line lists for computing the cross sections. The H$_2$O and CO line lists are obtained from the HITEMP database \citep{rothman_2010} and that of HCN from the Exomol database \citep{harris_2006,barber_2014,tennyson_2016}. The individual lines are broadened by both temperature and pressure using the Voigt profile following the methods in \cite{gandhi_2017}. These line lists contain $\sim$750,000 transitions for HCN in the 3.15-3.30$\mu$m, and the 2.25-2.37$\mu$m range contains $\sim$1.6$\times$10$^6$ transitions for H$_2$O and $\sim$2,800 transitions for CO. The molecular cross section for each species is generated by combining the contribution from all lines on a 0.01 cm$^{-1}$ spacing wavenumber grid in both spectral ranges, corresponding to $R\sim$300,000 to 400,000. We consider $\sim$10$^2$-10$^3$ spectral lines in our cross-correlation templates, optimized for each species that we consider.

We perform a nested analysis varying the number of lines in the spectral templates used for the cross correlation to determine the number required for a strong signal. We choose the strongest spectral features to generate these spectral templates. We find that a minimum of $\sim$250 lines are required to obtain the HCN S/N of $>4$. The peak S/N is obtained with $\sim$750 spectral lines. Beyond this, the signal does not increase as weak lines do not strengthen the cross correlation.

Figure \ref{fig:flux} shows the planetary flux for both spectral ranges for H$_2$O, CO, and HCN. As discussed in section~\ref{sec:results}, the  peak cross correlations were made using volume mixing ratios of CO $=10^{-3}$, H$_2$O $=10^{-4}$ and HCN $=10^{-5}$. The CO spectrum in the 2.3$\mu$m range shows deep absorption features resulting from the prominent lines and the upwardly decreasing temperature profile. The H$_2$O shows weaker features owing to its lower atmospheric abundance and smaller cross section. Thus, despite the significantly greater number of H$_2$O transitions, CO yields a higher S/N in this spectral range (Figure \ref{fig:test}). The 3.2$\mu$m range also shows significant absorption features in the spectrum arising from the numerous strong HCN lines. Despite the lower abundance of HCN we are able to obtain strong features in the spectrum at this wavelength thanks to its large molecular cross section. 

\section{Results and Discussion}\label{sec:results}

We find evidence for the presence of the species CO, H$_2$O, and HCN in the dayside atmosphere of HD 209458b. The cross-correlation S/N for each molecule over a grid of $K_p$ and $V_{sys}$ are shown in Figure \ref{fig:test}. In the 2.3$\mu$m window, cross correlation with H$_2$O and CO templates yield S/Ns of 4.4 ($K_p$ = 143$^{+15}_{-13}$ km s$^{-1}$ and $V_{sys}$ = -14$^{+6}_{-5}$ km s$^{-1}$) and 5.1 ($K_p$ = 136$^{+15}_{-14}$ km s$^{-1}$ and $V_{sys}$ = -16$^{+6}_{-5}$ km s$^{-1}$), respectively. Using a combined H$_2$O + CO model gives a boosted 5.3 S/N at $K_p$ = 141$^{+11}_{-14}$ km s$^{-1}$ and $V_{sys}$ = -15$^{+4}_{-4}$ km s$^{-1}$. Cross correlation with an HCN template yields a peak S/N of 4.7 at $K_p$ = 142$^{+21}_{-13}$ km s$^{-1}$ and $V_{sys}$ = -14$^{+5}_{-7}$ km s$^{-1}$ in the 3.2$\mu$m spectral window. The cross-correlation signal for HCN is made possible by its high molecular opacity in the 3.2$\mu$m band, and a dense forest of deep absorption lines. The presence of CO is consistent with previous detections, both in transmission \citep{snellen2010} and emission \citep{schwarz2015, brogi2017}.

The $K_p$ and $V_{sys}$ of the detections are in precise agreement with those determined in previous studies $K_p$ = 140$^{+10}_{-10}$ km s$^{-1}$ and $V_{sys}$ = -14.8 km s$^{-1}$ from orbital parameters as well as previous studies \citep{snellen2010}. 

\begin{figure}
\centering
  \centering
  \includegraphics[width=.49\linewidth]{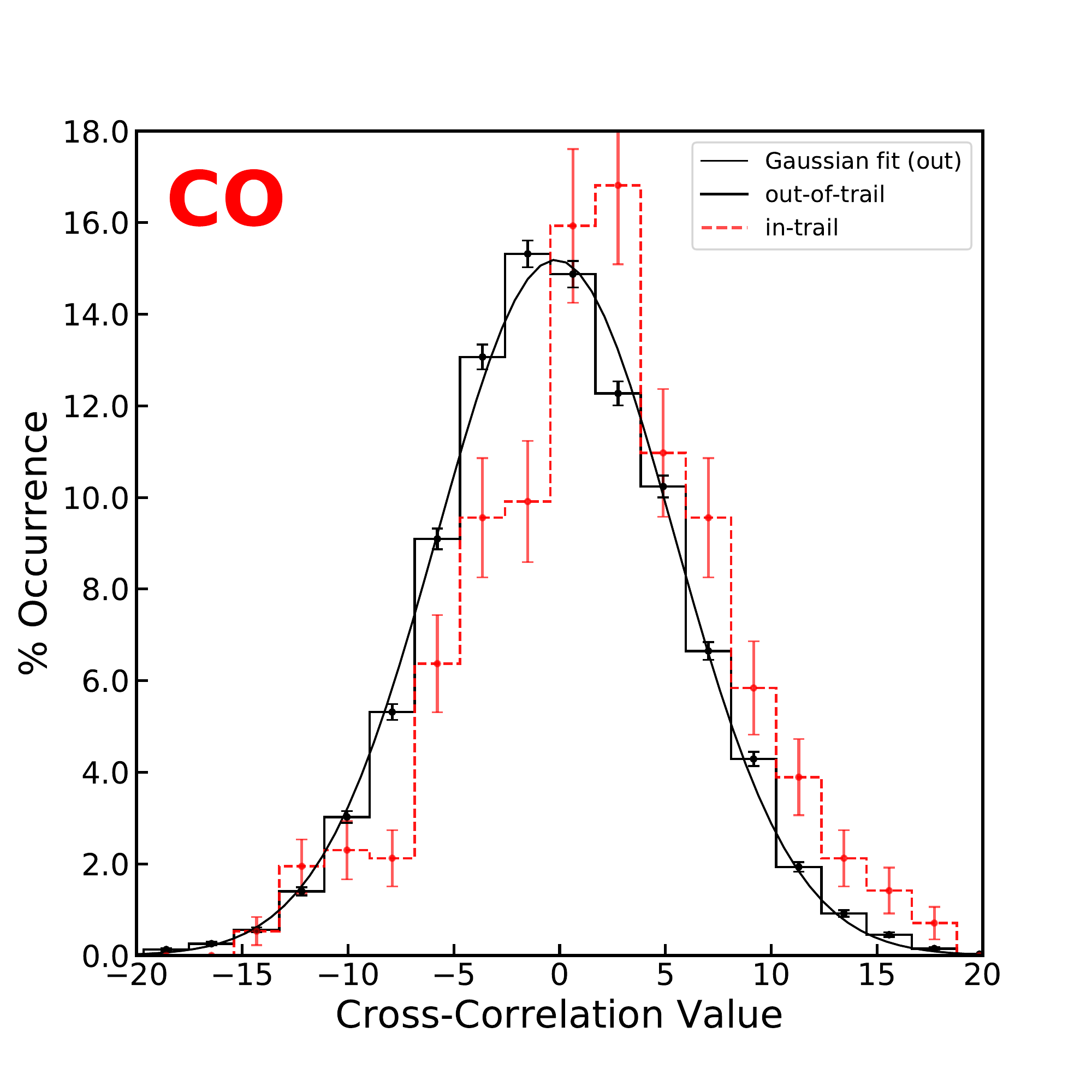}
  \includegraphics[width=.49\linewidth]{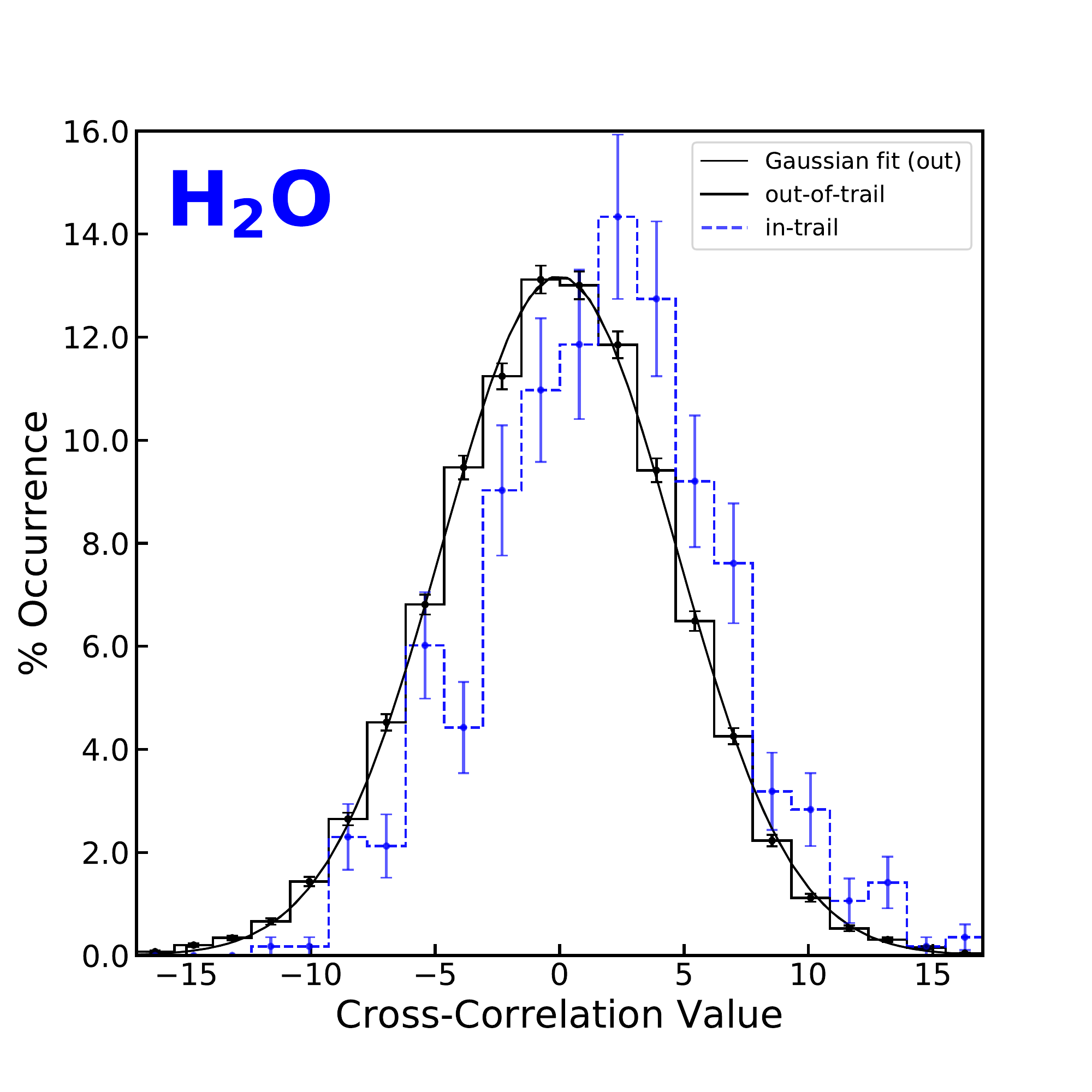}
  \centering
  \includegraphics[width=.49\linewidth]{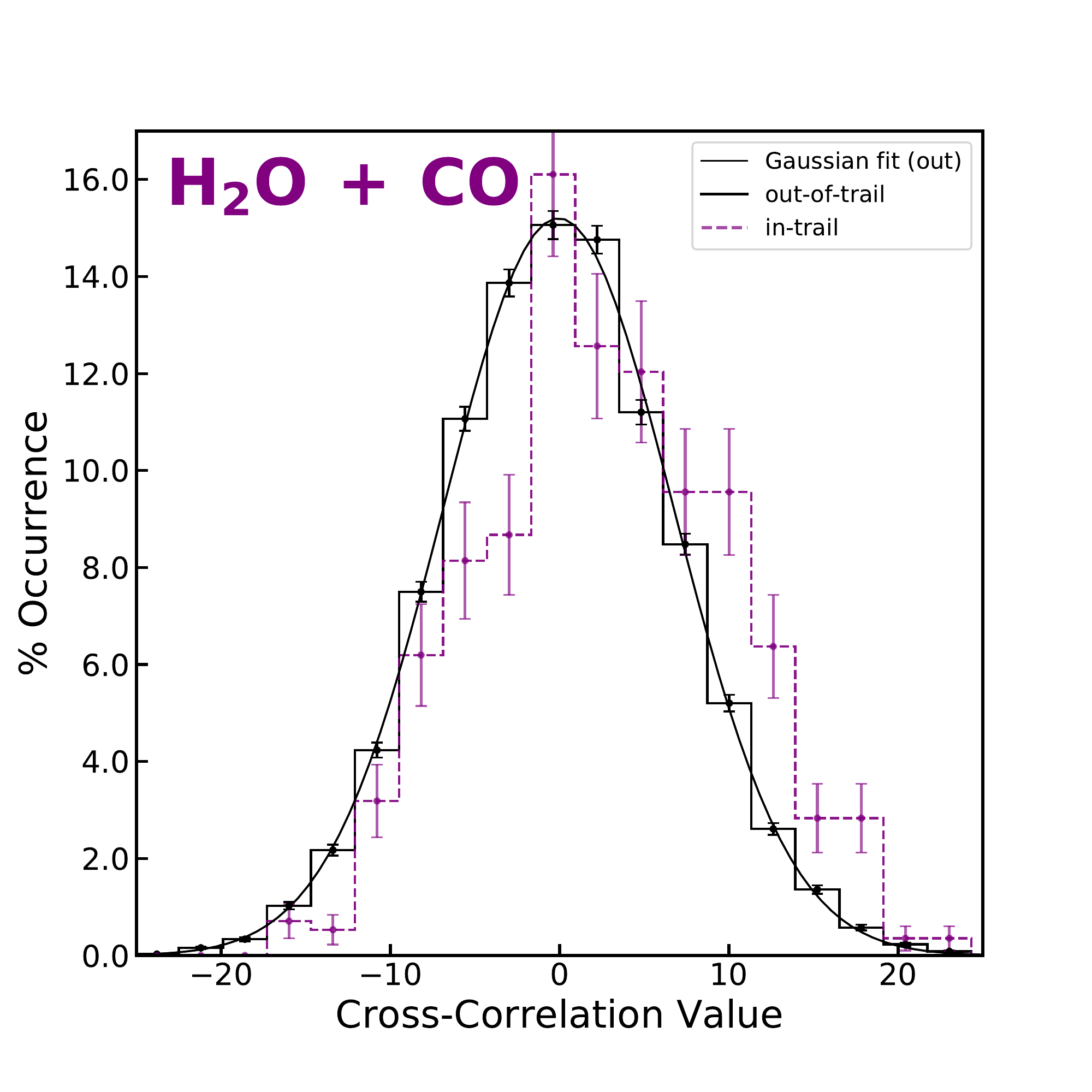}
  \includegraphics[width=.49\linewidth]{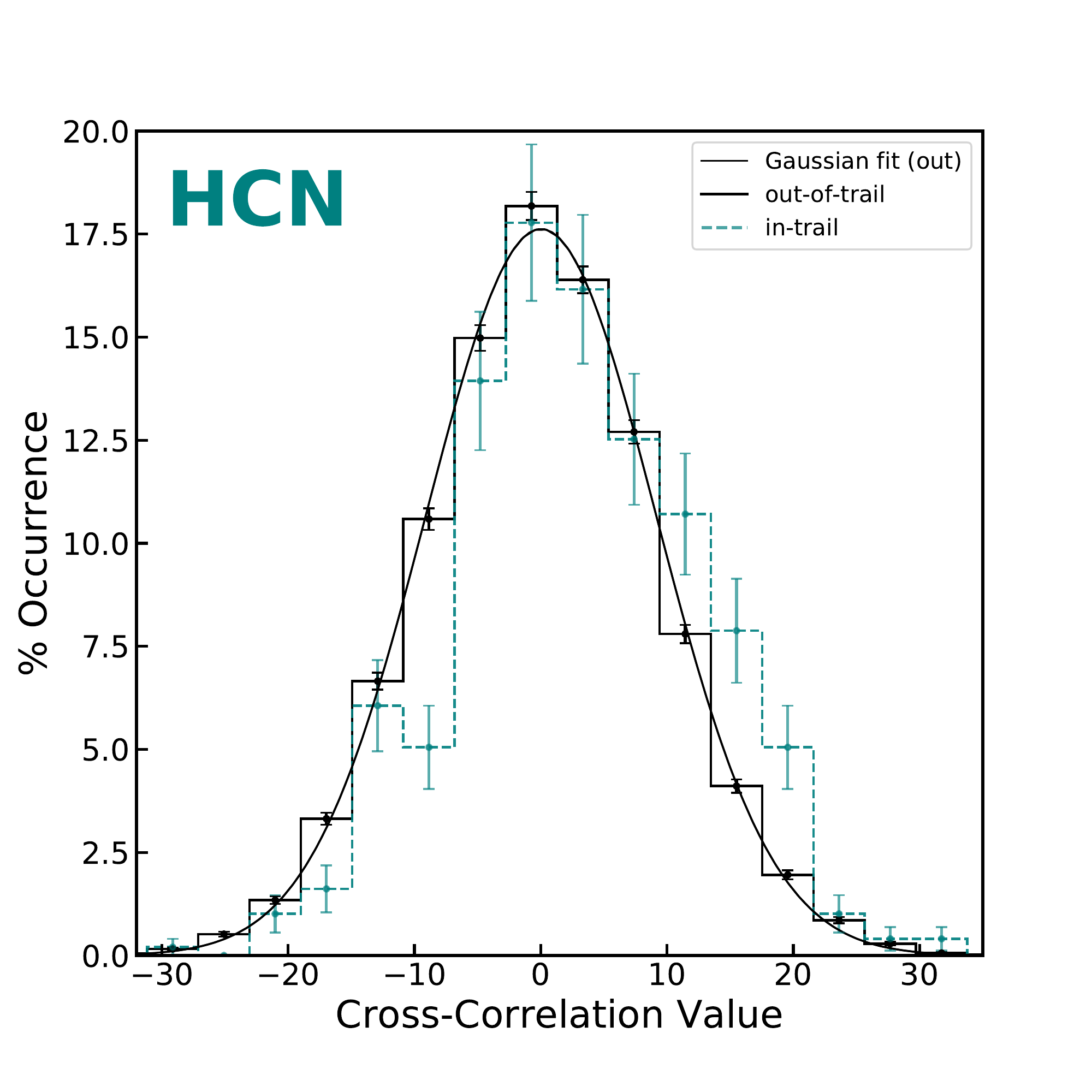}
\caption{
In-trail and out-of-trail samples of the cross-correlation function for each molecular detection. Note the consistent shift of the in-trail samples toward higher means and the strong Gaussian nature of the out-of-trail samples. Welch $T$-tests suggest that the samples are drawn from different distributions with greater than 5.7$\sigma$ confidence for all cases.}
\label{fig:in_vs_out}
\end{figure}

The S/N provides a metric for how well the model template matches intrinsic features. However, the cross correlation is sensitive to the stages of detrending outlined in section~\ref{sec:obs}, and the choice of model template. SYSREM does not exactly remove systematics and environmental effects, but rather approximations of them that best match a linear trend in the data. These approximations depend on the choice of columns that are masked. For example, masking less data around tellurics increased the number of spectral features with which we cross correlate, but requires more SYSREM iterations to fully remove higher-order residuals, which in turn can degrade the intrinsic planet signal. In optimizing the number of SYSREM iterations using model-injection, a commonly accepted method \citep{birkby2013,birkby2017,nugroho2017}, one potentially introduces the possibility of `overfitting' the detrending hyperparameters and obtaining a higher S/N than truly reflected within the data, especially given the fact that each detector is treated separately. We perform a series of tests, concentrating on the 3.2$\mu$m data, in which we (1) change the amount of masking around telluric zones; (2) optimize and search for other models including H$_2$O, CH$_4$, NH$_3$, and ones where the lines are randomly shuffled; and (3) optimize the number of SYSREM iterations to other local maxima in $K_p$-$V_{sys}$ space. 

As expected, varying the amount of masking has a significant effect on the detection significances and optimum number of SYSREM iterations. HCN is the only model with which we obtain a robust cross-correlation S/N$>4$ at the planetary $K_p$ and $V_{sys}$ in the 3.2$\mu$m band. However, when optimizing to other local maxima in $K_p$-$V_{sys}$ space, in some cases we are able to obtain cross-correlation S/N values $\geq 4$ at locations inconsistent with the expected $K_p$ and $V_{sys}$ of the planet. This raises some concern about potential false positives, especially if such a signal should happen to coincide with the known $K_p$ and $V_{sys}$ or if $K_p$ and $V_{sys}$ are not known from another method. It could be that the variation of hyperparameters effectively allow enough draws to be taken from the noise distribution, thus giving an S/N of $\geq 4$. We find that the Welch $T$-test metric commonly used in the literature \citep{birkby2017,nugroho2017} is similarly vulnerable and returns even higher estimates of the signal significances (see Figure 5), hence we choose to use the more conservative S/N metric. An ideal statistical metric for detection significances should account for false positives and the sensitivities inherent to the method.  

We perform an independent test using the detrending approach detailed in \cite{schwarz2015} and \cite{brogi2014}. This detrending approach is not generally used for data in the 3.2$\mu$m spectral window as SYSREM is preferred \citep{birkby2013,birkby2017,birkby2018}, which tends to be better at removing the stronger tellurics at these wavelengths. Nevertheless, we apply the alternative detrending method masking only a small percentage of the data on the ends of the detectors. We then linearly fit the logarithmic flux of each column with airmass variation, and remove the trend by division. Next, we sample columns from deep telluric features and linearly regress them to each column. We divide by the fit, which removes the second-order residuals. Finally, we normalize each column by its variance to suppress noisy pixels. Unlike the SYSREM approach, this method does not involve optimization, and reduces the chance of overfitting. However, we are unable to obtain an S/N $\geq 3$ from cross correlation with the HCN template in the 3.2$\mu$m band. This may be partly due to the presence of remaining systematics which SYSREM is more effective at removing. The best method to remove telluric contamination is yet to be determined \citep{birkby2018}. We find that, applied to the 2.3$\mu$m, the alternative detrending method produces a similarly inconclusive ($\sim3\sigma$) potential CO detection in agreement with \cite{schwarz2015}; it also suggests that it is indeed worse at removing telluric systematics.

We consider a range of abundances when performing the cross correlation. We obtain the strongest CO signal using a mixing ratio of 10$^{-3}$ based on previous high-resolution analysis \citep{snellen2010}. We obtain the H$_2$O signal using a mixing ratio of 10$^{-4}$ that is consistent with previous estimates using low-resolution spectra of HD~209458b \citep{madhu2014b,brogi2017,macdonald2017}. Given no previous detection of HCN, we explore a range of mixing ratios between $\log$(HCN) of -8.0 and -3.0, and subsequently use a best-fit mixing ratio of 10$^{-5}$. We also obtain a lower bound on the atmospheric abundance of HCN in the planet for obtaining an HCN signal with a S/N $\geq3$. For this purpose, we first subtract the best-fit model planetary signal from the data, which reduces the cross-correlation detection significance to zero. We subsequently inject an artificial planetary signal corresponding to model spectra for each explored abundance, and try to detect it by cross-correlating with the same model. We find that a minimum $\log$(HCN) of -6.5 in the planet is required to obtain a SNR of $\geq3$. If confirmed, such a minimum atmospheric abundance of HCN could imply a high C/O ratio ($\sim$1 or higher) in the dayside atmosphere of the planet \citep{madhu2011,madhu2012,moses2013}. 

\acknowledgments
N.M., G.H., and S.G acknowledge support from the UK Science and Technology Facilities Council (STFC). 

\vspace{5mm}

\bibliography{hires}
\bibliographystyle{yahapj}
\end{document}